cond-mat/9409054

-------------------------------------------------------------------------
\documentstyle[preprint,aps]{revtex}
\begin{document}
\draft

\title{Spectral properties of classical two-dimensional clusters.}

\author
{Vitaly A. Schweigert \cite {*:gnu} and  Fran\c cois M. Peeters \cite {f:gnu}}

\address{\it  Departement Natuurkunde, Universiteit Antwerpen (UIA),\\
Universiteitsplein 1, B-2610 Antwerpen, Belgium}

\date{\today}
\maketitle
\begin{abstract}

	We present a study of the spectral properties like the
energy spectrum, the eigenmodes and density of states
of a classical finite system of two-dimensional (2D)
charged particles which are confined by a quadratic potential.
Using the method of Newton optimization we obtain the ground state and
the metastable states. For a given configuration
the eigenvectors and eigenfrequencies for the normal modes  are
obtained using the Householder diagonalization technique
for the dynamical matrix whose elements are the second
derivative of the potential energy.
For small clusters the lowest excitation corresponds to an
intershell rotation. The energy barrier for such
rotations is calculated. For large clusters the
lowest excitation consists of a vortex/anti-vortex pair.
The Lindeman melting criterion is used to calculate
the order-disorder transition temperature for
intershell rotation and intershell diffusion.
The value of the transition temperature at which intershell rotation
becomes possible depends very much on the configuration of the cluster,
i.e. the distribution of the particles between the different shells.
Magic numbers are associated to clusters which are most stable
against intershell rotation.
The specific heat of the  cluster is also calculated using
the Monte-Carlo technique which we compare with an analytical
calculation where effects due to anharmonicity are incorporated.
\end{abstract}

\pacs{PACS numbers: 64.60.Cn, 64.70.Dv, 73.20.Dx}

\section { Introduction}
	During the last few years  considerable attention has been paid to
the  study   of the properties of mesoscopic systems consisting
of a finite number of neutral or charged particles.
The particles are confined  by an artificial external confining field.
Behavior of either ions in a radio-frequency trap (Paul trap) or
a Penning trap \cite {11,42}
and heavy-ion ring storage \cite {12} can serve as
an illustration of three-dimensional (3D) Coulomb clusters.
Very large Coulomb clusters have been created recently in strongly
coupled rf dusty plasmas \cite{40}.
Examples of two-dimensional (2D) Coulomb
clusters are electrons on the surface of liquid He \cite {13}
and electrons in quantum dots \cite {14}.
The vortex clusters in an isotropic superfluid \cite  {15} and in
superconducting grains \cite{ree} have
many common features with those of 2D charged particles \cite {16}.
Refs. \cite {17,18} have been devoted to the investigation of the
ground state of 3D clusters of charged particles.
Below we give a short overview of previous theoretical work on
2D clusters of charged particles.

	Clusters of particles in 2D with Coulomb repulsion were
investigated by Lozovik and co-workers \cite {2}
in the case of parabolic confinement. They found that for
low temperature and in the case of a small number of particles
the cluster has a shell structure. A two step order-disorder transition
was found. With increasing
temperature, first intershell rotation starts, and
intershell diffusion may be possible at high temperature.
When the size of the cluster is sufficiently large,
the  simple shell structure gradually disappears in the  center
and features of a Wigner lattice appear.
Then cluster melting occurs around the 2D Wigner lattice melting temperature.

	Bolton and R\" {o}ssler \cite {6} considered
the case of parabolic confinement for a small number
of particles: $1-40$. They investigated the ground state as well as some
metastable states. For clusters consisting of 6 particles they
determined the barrier height for transition from the configuration
(1,5) (these are the number of electrons in each shell)
to the configuration (6).

	Systematic and detailed investigation of the structure of 2D
clusters was carried out by Bedanov and Peeters \cite {5}. They
considered both parabolic and hard wall confinement. A table  of
Mendeleev was constructed for clusters with: $2-52$, $82$, $151$, $230$
number of particles.
Using the Lindeman melting criterion these authors determined the
temperature for the order-disorder transition for radial and
angular displacement.

	In all of the above works on 2D systems with a finite
number of charged particles the Monte-Carlo simulation technique was used.
We found that in some cases this method is rather slow in finding the
ground state of the cluster. The reason is that
the Monte-Carlo technique spends too much time in the vicinity of
metastable states such that for a finite simulation time not necessarily
the correct ground state is found. This becomes more of a problem for
clusters with larger number of particles which have many more metastable
states. In Ref.\cite {5} this drawback was partially avoided by
heating up the system and cooling it down repeatedly.
In the present work we will present an alternative approach. To find the
ground state we choose the Newton method with initial
configurations determined randomly. In this way we are able to obtain
not only the ground state but also the metastable states. The latter
are relevant in the calculation of thermodynamic properties
and the barrier height for intershell rotation.

	In previous work \cite{2,6,5} the ground state properties
and melting temperatures were obtained.
Here we will investigate the spectral properties
of the system. This paper is organized as follows.
In  Sec. II we describe the model and
introduce the dimensionless units. In Sec. III our numerical technique
to obtain the ground and metastable states is outlined and compared to the
Monte-Carlo technique. Sec. IV is devoted to the stable configurations
and the spectrum of normal modes is determined.
The barrier height for intershell rotation is obtained in Sec. V.
Intershell rotation is the lowest excitation for small clusters.
We correlate the strong dependence of the height of the barrier for
intershell rotation to the number of particles placed in various shells.
In Sect. VI we discuss large clusters for which we calculate the density
of states and discuss their lowest excitation which consists of a
vortex/anti-vortex pair.
In Sec. VII the zero temperature results for the excitation spectrum
are used in order to calculate the melting temperatures using the Lindeman
melting criterion.
These results are compared with earlier results \cite{5} which
were based on the Monte Carlo simulation technique.
As an example of the use of metastable states in the calculation of
thermodynamic property we calculate the heat capacity in Sec. VIII. We compare
the Monte-Carlo results with an analytical approach in which we include
anharmonicity effects in an approximate way.
Our conclusions are presented in Sec. IX.

\section { Model system}
	The model system was defined in Ref. \cite {5}.
But for completeness we recall the main features.
Our system is described by the Hamiltonian
\begin{equation}
\label{eq1}
H=\frac {q^2}{\epsilon }
\sum_{i>j} \frac {1}{\mid \vec {r}_i-\vec {r}_j\mid}
+\sum_{i}V(\vec {r}_i),
\end{equation}
where $q$ is the particle charge, $\epsilon$ is the dielectric
constant of the medium the particles are moving in and the confinement
potential $V(\vec {r})=\frac{1}{2}m\omega_{0}^{2}r^2$ is taken
parabolic. Particle motion is described by classical mechanics in the plane
$\vec{r} = (x,y)$. To exhibit the scaling of the system we introduce the
characteristic scales in the problem:
$r_0=(2q^2/m\epsilon \omega _{0}^2)^{1/3}$ for the length,
$E_0=(m\omega _{0}^2q^4/2\epsilon^2)^{1/3}$ for the energy, and
$T_0=(m\omega _{0}^2q^4/2\epsilon^2)^{1/3}k_B^{-1}$ for the temperature.
These scales will be used as our new units and all our results will be
given in these units. In so doing the Hamiltonian can be written as
\begin{equation}
\label{eq5}
H=\sum_{i>j} \frac {1}{\mid \vec {r}_i-\vec {r}_j\mid}
+\sum_{i}V(\vec {r}_i) ,
\end{equation}
with $V(\vec {r})=x^2+y^2$. The numerical values for the
parameters $\omega _0$, $r_0$, $E_0$, $T_0$ for some typical
experimental systems were given in \cite {5}.

	In the present paper we will consider only classical
systems. Although a classical approach for the description
of the behavior of electrons in quantum  dots is not applicable,
nevertheless it is possible that certain features of the classical
system may survive in a quantum system. For example in the quantum study  of
the transition from a crystal to a liquid in the absence of a
magnetic field\cite {7}, we know that
the parameter for formation  of  a  Wigner
crystal is $r_s=l_{0}/a_0=37\pm5$,
where $l_0$ is the mean distance between the particles.
If the number of particles is small,
the interparticle distance in the case of parabolic confinement is close to
$r_0$. Thus for typical parameters for a
quantum dot in GaAs with $m=0.067$, $\epsilon =13$, $ \hbar \omega_0
= 1\: meV$ we obtain $r_s=7.8$. Reducing the confinement $\omega_0$ or
applying a magnetic field \cite{102} will give us
a possibility to investigate the existence of a Wigner
crystal or another ordered state for a finite number of particles.
In Ref. \cite{5} it was found that a classical 2D
cluster with a finite number of charged
particles can be more or less stable than a 2D
crystal for the same parameter $\Gamma=q^2/\epsilon l_0 k_BT$.
We expect that a similar quantity will be relevant in the
quantum case and therefore it is expected that
also a Wigner crystal like state can exist in quantum dots.

\section {Numerical approach}

	The Monte-Carlo simulation technique \cite{8} is relatively simple and
provides relatively rapid convergence and a reliable
estimate of the total energy of the system in cases that a relative
small number of Metropolis steps is sufficient.
However, the accuracy of this method in calculating the
explicit states may be poor in certain cases.
We can understand this as follows: for the present system of
axial symmetric confinement some configurations have very
small frequencies for intershell rotation
$\omega_{min}=10^{-3}\div 10^{-4}$ which may lock the simulation in an
unstable state.
Using the Monte-Carlo method with an
acceptable number of steps $10^4\div 10^5$, in order to limit the
computer time, we may obtain the energy $E$ up to an error $\delta $,
but the error in the coordinates will be proportional
to $\delta ^{1/2}/\omega_{min}$ which in such a case can be large.

	To circumvent this problem we used a different numerical approach
which is mainly based on our experience from which we learned that
with different modifications to the gradient method and
the method of molecular dynamics using artificial viscosity
we were able to obtain more reliable results than with
the Monte-Carlo technique.
To be more explicit, to find the state with the minimal energy we used
the modified Newton technique. Since this method is
practically not applied in the present field we will give a short outline.
Let us suppose that  the coordinates of the particles in a cluster are given
by \{$r_{\alpha ,i}^n$; $\alpha =x,y$, $i=1,\ldots N$\}
after $n$-steps in the simulation.
Then the potential energy in the vicinity of this configuration
can be written in the following quadratic form
\begin{equation}
\label{eq6}
	H=H(r_{\alpha,i}^n)
	-\sum_i \sum_{\alpha} H_{\alpha,i}(r_{\alpha,i}-r_{\alpha,i}^n)
	+\frac {1}{2} \sum_{i,j} \sum_{\alpha ,\beta} H_{\alpha \beta, \: ij}
	(r_{\alpha,i}-r_{\alpha,i}^n)(r_{\beta,j}-r_{\beta,j}^n) ,
\end{equation}
where $H_{\alpha,i}=-\partial H/\partial r_{\alpha,i}$ is the force
and $H_{\alpha \beta \:,ij}$ is the dynamical matrix
\begin{equation}
\label{eq7}
	H_{\alpha \beta , \:ij}=
	\frac {\partial ^2H}{\partial r_{\alpha,i}\partial r_{\beta,j}} .
\end {equation}
The next step in our simulation is based on the condition
of minimal total energy
\begin {equation}
\label {eq8}
	\sum_j \sum_{\beta} (\eta \delta_{\alpha \beta ,\: ij}
	+H_{\alpha \beta \:,ij})
	(r_{\beta,j}-r_{\beta,j}^n)=H_{\alpha,i} ,
\end {equation}
where $\delta _{\alpha \beta ,\: ij}$ is the unit matrix and the coefficient
$\eta $ is added to assure the stability  of
the algorithm. It is easy to show that the iteration
procedure  converges if $\eta > -\lambda _{min}$, where $\lambda_{min}$ is
the minimal eigenvalue of the dynamical matrix.
The system of linear equations (\ref {eq8})
is solved  using Gaussian elimination.
The calculation of the matrix and solving the system  of  linear  equations
takes about $N^2$ numerical operations.
This is equivalent to a Monte-Carlo step where also about $N^2$
operations are needed to find the energy, but the
coefficient in front of $N^2$ is less for the latter. The reason is that
to obtain the spectrum of the matrix is more laborious. The usual approach
guarantees only convergence in the vicinity
of the minimum. Therefore we introduced an empirical dumping coefficient
$\eta $. In the first few iterations the value for $\eta $
is set to be large: $\eta =10\div 100$. If in the next step
the total energy of the system decreases the dumping coefficient
is reduced, while in the opposite case the value $\eta$ is
increased. From our experience we know that such
an algorithm for choosing the dumping parameter guarantees
convergency of the iteration process. Furthermore, near the last steps,
the dumping parameter becomes less than the minimal
value of the eigenvalue of the dynamical matrix and the rate of convergency
becomes square ( $\delta_{n+1}\sim \delta_n^2$).
The accuracy of the calculated energy $\delta $
is now only limited by rounding errors. For systems with axial
symmetry there exists an eigenvalue with value zero which corresponds
to turning the system as a whole around the axis of symmetry.
In such a case the second eigenvalue $\lambda_2$ has to be
taken as the minimal eigenvalue.
We found that in order to obtain the configuration with minimal energy with an
accuracy of $\mid H_{\alpha,i}\mid =10^{-9}\div 10^{-10}$
takes about $10 \div 100$ steps, the exact number of steps
depends on the number of particles.

	After finding the state with the minimal energy
we obtain the eigenvalues and eigenvectors of the dynamical matrix (\ref
{eq7}).
The eigenfrequencies of it are the eigenvalues squared.
The condition that the minimal eigenvalue is positive
guarantees that the obtained configuration is stable.
Of course also the present method does not guarantee that all stable
and metastable configurations and the configuration with the lowest energy
are found.
To overcome this difficulty partially we consider a large number (typically
$200$) of initial configurations which are generated randomly.
{}From these initial configurations a few stable
configurations remain, the number of which increases fast
when $N>30\div 40$. Among these stable configurations
the state with the lowest energy is taken to be the ground state
of the system. The fact, that usually the state with the minimal
energy is achieved already after a small number of steps, gives us confidence
that this is likely the actual ground state of the cluster.
Usually, the radius of convergence of the ground state is sufficiently large.
We confirmed that the present approach for $N<80$ leads to the ground state
configurations of Ref. \cite{5} which were obtained using the Monte-Carlo
method with about $10^5 \div 10^6$ simulation steps.

	The efficiency of the present method is illustrated in Fig.1
where we plot the precision of the energy, which is defined as the
difference from the exact energy value, as function of the number of
simulation steps for a cluster of $13$ particles.
It is apparent that the present technique converges much faster, about
an increase with a factor of $200$ is found.
Furthemore, we discovered that even if within the Monte-Carlo approach the
error in the energy is only of order
$10^{-11}$, the obtained cluster configuration was unstable. This was found
by calculating the minimal eigenvalue of the matrix which consists of
the second derivative of the potential energy
with respect to the position coordinates which for the obtained configuration
was negative.
The present Newton optimization approach did not exhibit such a deficiency.
In contrast to the Monte-Carlo approach of Bolton and
R\" {o}ssler \cite{6} who found more than one stable configuration for
the case of $N=13$ particles, the present approach in which $200$ initial
configurations were considered, demonstrates that there exist only one stable
configuration which is (4,9). But for this configuration the minimal
excitation frequency $\omega _{min}\approx 6\cdot 10^{-4}$
is very small which may be the reason for the error in Ref. \cite{6}.

\section {Eigenvalues and eigenvectors}

	A detailed  description of the features of the lattice structure,
the interparticle distance
scale in the various shells, and the Mendeleev table for the configurations
with $N=2-52$, $82$, $151$, $230$ particles was given in Ref. \cite {5}.
Here, we will discuss the excitation spectrum corresponding to the
ground state configuration of the system. This spectrum is shown
in Fig. 2 as function of the number of particles for $N$ ranging from 2 to 50.
The eigenfrequency in this figure is in units of $\omega_o/\sqrt{2}$.
Notice that there are three eigenfrequencies
which are independent of N: i) for any axial symmetric system the system
as a whole can rotate which gives an eigenfrequency
$\omega=0$. This is illustrated in Fig. 3 (figure indicated by $k=1$; $k$
counts
the eigenvalues in increasing order) where the arrows
indicate the direction of movement of the different particles (i.e. the
eigenvectors of the excitation) for a cluster with $N=9$; ii) there is a
twofold degenerate vibration of the center of mass with frequency
$\omega =\sqrt {2}=1.4142$ (see Fig. 3, $k=7$ ); and iii) the
third eigenfrequency corresponds to a vibration of the mean square radius
$N^{-1}\sum _{i}(x_i^2+y_i^2)$ with frequency $\omega=\sqrt {6}=2.4495$
(see Fig. 3, $k=15$). The value of this breathing mode
can easily be obtained analytically.

	For clusters of sufficient large size (i.e. $N>8$)
a typical feature of its spectrum
is the occurrence of a very low eigenfrequency. Because of the scale in
Fig. 2 this frequency is not discernable from the $\omega=0$ frequency
and therefore we have listed it in Table I as $\omega_{min}$.
For a number of clusters the eigenvectors corresponding to these minimal
eigenvalues are shown in Figs. 3($k=2$ ), 4 and 5($k=2$ ).
For $N=19$ and $N=20$ the central particle does not move for this
specific excitation and consequently its displacement vector has length
zero and is therefore not visible in Fig. 4.
For the clusters with $N=9, 19$ and $20$ particles,
the motion with the minimal frequency
$\omega_{min}$ corresponds to intershell rotation.
The necessary condition for the existence of intershell rotation is
the presence of at least two particles on the inner shell in order
to conserve total angular momentum. With changing configuration, the minimal
eigenfrequency can vary by several orders of magnitude (see Table I).
For instance, for $N=19$ with the ground state configuration ($1,6,12$),
the minimal eigenfrequency is $\omega _{min}\approx0.67$,
and for $N=20$ and configuration ($1,7,12$),
$\omega _{min}\approx1.0\times10^{-4}$. In both cases the minimal
eigenfrequencies correspond to intershell rotation (Fig. 4).
This large change in the size of
the minimal eigenfrequency is connected with the shell configuration,
and not with the total number of particles.
For example, if for $N=19$ we take the metastable configuration
($1,7,11$) whose energy is an amount
$1.66\times10^{-2}$ larger than the ground state energy, we obtain
$\omega _{min}\approx1.1\times10^{-4}$ which
coincides practically with $\omega _{min}$ for the cluster
with $20$ particles.

	From the data given in Table I we infer the following
law: {\it a high frequency value for
intershell rotation is obtained for configurations such that the number
of particles on the outer shell is an integral
number times the number of particles on the inner shell}. For example:
$N=12$ ($3,9$), $N=15$ ($5,10$), $N=16$ ($1,5,10$), and $N=19$  ($1,6,12$).
For clusters with more than two shells (i.e. $N>21$) a large
$\omega_{min}$ for intershell rotation is found for ground state configurations
in which the number of particles in the different shells are multiples of an
integer number. The latter is usually the number of particles in the
inner shell. For example: $N=22$ ($2,8,12$), $N=30$ ($5,10,15$),
$N=45$ ($3,9,15,18$) and to a lesser extend also $N=34$ ($1,6,12,15$).
These cluster numbers can be considered as the {\it magic numbers}, because
they represent the clusters which are most stable against intershell
rotations. In previous work by others on 3D clusters magic numbers
were determined
on the basis of energy calculations of the cluster configuration.
We found \cite{5} that for 2D clusters no clear steps are found in the
cluster energy versus the number of particles in the cluster and therefore
the stability argument is more appropriate in the present case.
On the other hand, a configuration with small $\omega_{min}$ for
intershell rotation is realized when the number of
particles in the different shells have no common denominator.
For example: $N=13$ ($4,9$) and $N=20$ ($1,7,12$).

	The above rules can be understood from the Hamiltonian by analyzing
the intershell interactions using cylindrical coordinates.
Let us consider the most simple configuration with two shells.
{}From the outset we notice that the occurrence of an particle in the
center of the system,
for example, for a cluster with $20$ particles, does  not  disturb
the intershell rotation. Therefore, we do not have to
consider such a case separately. Let us discuss the rotation between two
outer shells.
The interparticle distance and the  distance  between  the  shells
changes only by a few percent when we alter
the number of particles and/or the configuration.
Therefore, in an initial approximation we can describe  each shell
by an ideal polygon and thus the interaction Hamiltonian between
two shells can be reduced to the form
\begin {equation}
\label {eq9}
H=\frac {1}{2} \sum _{i=1}^{N_1}\sum _{j=1}^{N_2}
(R_1^2+R_2^2+2R_1R_2\cos {(i\theta _1-j\theta _2-\theta)})^{-1/2} ,
\end {equation}
where $R_1$, $R_2$, $\theta_1=2\pi/N_1$, $\theta_2=2\pi/N_2$
are the radii and angles between particles of the first and second
shell which have $N_1$ and $N_2$ particles respectively, and $\theta $
is the intershell angular distance. The sum (\ref {eq9}) over the two indexes
can be reduced to the sum over one index only
\begin {equation}
\label {eq10}
	H=\frac {N_1N_2}{2K} \sum_{i=1}^{I}
	(R_1^2+R_2^2+2R_1R_2\cos {(i\theta _{\star }-\theta )})^{-1/2},
\end {equation}
where
\begin {equation}
\label {eq11}
	\theta_{\star}=\frac {2\pi} {I},
\end {equation}
and $I$ is an integer which is the minimal divider of the number of
particles ($N_1$, $N_2$, ...) in the different shells.
{}From expression (\ref {eq10}) it follows that the Hamiltonian for intershell
interaction is periodic in $\theta$.
Moreover the period $\theta _{\star}$, as a rule is less
than the angular interparticle distance within a shell.
To evaluate the strength of the intershell interaction we deduct from
the Hamiltonian the following value
\begin {equation}
\label {eq112}
	\frac {N_1N_2}{4\pi } \int _{0}^{2\pi} dx
	(R_1^2+R_2^2+2R_1R_2\cos { (x-\theta) })^{-1/2} ,
\end {equation}
which is independent of $\theta$. This result was
obtained from Eq. (\ref{eq10}) by replacing the summation over $k$ by an
integration. We proved that the error we make in doing so is
proportional to $\theta_{\star }^2$. Numerical summation of ($\ref{eq10}$)
gives even a more weaker dependence of the interaction energy on $N$ for two
ideal polygons. When we compare the computed results for the barrier height for
intershell rotation with those found from Eq.(9) we found that
Eq.(9) gives a good qualitative description but
quantitatively the results are not satisfactory.
Therefore we may conclude that for small eigenvalues,
the exact value of the barrier height is strongly influenced by the
{\it non-ideality} of the polygons. Indeed in order to obtain Eq.(6) we assumed
that the particles were placed at the edges of an ideal polygon.
Because intershell rotation is a collective phenomena,
one can easily understand that
the actual barrier height is less than that given by Eq. (\ref {eq10})
due to the deformation of the polygons during the motion.
Indeed, during the rotational motion not only the intershell distance
changes but also the interparticle distance within a shell is altered.
This is illustrated in Figs. 3 and 4. From these figures we notice that
the eigenvectors for the particles in the inner shell have practically
the same length and are orthogonal to the radius-vector
of the particle. For the outer shell the situation is different
and the eigenvectors have also components in the radial
direction and futhermore, the length of the eigenvectors are
different for the different particles.
Therefore the vibrations in the radial and axial directions of the outer shell
follow the intershell rotational motion of the particles.
Only for clusters in which the number of particle on the inner shell is a
multiplicative integer factor of those of the outer shell, i.e.
when a large intershell rotation frequency is found,
are the polygons almost ideal which can be understood from symmetry
reasons and from our numericl results.
The characteristics and modelling of the intershell rotation will be given
in next section.

	When we increase the number of particles, we found that
for $N=39$ the motion with the minimal
eigenfrequency no longer corresponds to intershell rotation,
but rather consists of rotation of different individual regions of the cluster.
For $N\geq 60$ (see Fig. 4) the  rotation of an inner shell is
followed by the rotation
of individual polygons at the periphery of the cluster.
For $N\geq115$ we found that the minimal frequency $\omega_{min}$ no
longer corresponds to intershell rotation but corresponds to the excitation
of a vortex/anti-vortex pair (see Fig. 5 for N=151). Higher excitations
(see Fig. 5 with $k=4$ and $k=6$) may consist of multiples of such pairs.
In case of a cluster of $N=151$ particles the $7^{th}$ lowest excitation
corresponds now to an intershell rotation.
A more detailed discussion on the nature of low energy excitations of large
clusters is postponed to Sect. VI.

\section {Barrier for intershell rotation }

	In the present section we give a more detailed discussion of the
lowest non-trivial excitation in case of small clusters, which is the
intershell rotation.
The barrier for intershell rotation was obtained using the following
procedure. Let us assume that after n-steps in our simulation
the coordinates of the  particles are given by $\{\vec{r}_i^n; i=1,...N\}$.
After diagonalizing the dynamical matrix $H_{\alpha \beta \:,ij}$
we obtain the eigenvectors $\vec{A}_i(k)$ and the eigenvalues $\lambda_k
=\omega_k$. The particle coordinates for a next time step is then given by
\begin{equation}
	\vec{r}_i^{n+1}=\vec{r}_i^n+\sum_{k=2}^{2N}\tau_k\vec{A}_i(k)\ .
\end{equation}
Denote $\tau=\tau_{k^*}$ as being the lowest frequency for intershell
rotation which is taken to be constant and which sets the size of the
time step. The values of all other coefficients $\tau_k$
are found from the condition of minimal  potential energy.
This is done as follows: substitute the above expression in Eq.(3)
which gives us the total energy for the next step
\begin{equation}
	H_{n+1}=H_{n}
	+\sum_{i=1}^{N}\sum_{k=2}^{2N} \tau_k
	\frac{\partial H}{\partial\vec{r}_i} \cdot \vec{A}_i(k)
	+\frac{1}{2} \sum_{k=2}^{2N} \lambda_k\tau_k^2 \ ,
\end{equation}
from which we readily find the coefficients
\begin{equation}
 	\tau_k=-\lambda_k^{-1} \sum_{i=1}^{N} \frac{\partial
	H}{\partial \vec{r}_i}\cdot \vec{A}_i(k)\ .
\end{equation}

	Trajectories of the particles in four different clusters are depicted
in Fig. 6. As was mentioned above,
intershell rotation takes place in conjuction with radial oscillations.
The latter are more noticeable for clusters with high symmetry,
which have a relative large
frequency and consequently large barrier heights for intershell rotation.
In clusters with only two shells, particles in different shells rotate
in opposite direction in order to conserve total angular
momentum. Such a motion is defined completely by the angle of rotation
$\varphi$ of one shell relative to the second. When
there are three shells or more it is convenient to introduce the angle
of rotation $\varphi $ of the shell with the maximum angular velocity as an
independent parameter.
Fig. 7 illustrates the dependence of the potential energy on this parameter
$\varphi$ for two different clusters with $N=9$ and $N=40$ particles.
In general this function is well approximated by the simple relation
\begin {equation}
\label {eq15}
	U=\frac {U_{\star}}{2}\left[1-\cos{(\frac{2\pi\varphi}
		{\varphi_{\star }})}\right] ,
\end {equation}
where $U_{\star }$, and $\varphi _{\star }$ are the barrier height and the
period, respectively. The values for $U_{\star}$ and $\varphi_{\star}$
are given in Table I.
The above procedure is not able to give the value of the barrier height for
the clusters $N=39,42,47,51$ and $70$. The reason is that for those clusters
the minimal eigenvalue does not correspond to intershell rotation.
Notice that the cluster with $N=40$ has two minima in the potential energy.
One of this minima corresponds to a metastable state.
Of course $U_{\star}$ and $\varphi_{\star}$ in Table I are determined by
the global minimum and maximum. In a few cases,
the maxima in potential energy are sharper than that given by expression
(\ref {eq15}). The reason is that the energy for clusters with three
and more shells are not only a function of the angular position of the shell.

	For clusters with two shells, the parameter
$\varphi $ characterizes the motion of the inner shell. The angle of
rotation of the outer shell relatively to the inner one can be
obtained from the condition of zero total momentum
\begin {equation}
\label {eq16}
	\theta =(1+\frac {N_1R_1^2}{N_2R_2^2})\varphi\ ,
\end {equation}
where $N_i, R_i$ are the number of particles and the radius of shell $i$,
respectively. For clusters with two shells, the value $\varphi _{\star}$
presented in Table I is
correctly approximated by the simple analytical formulas (\ref {eq11})
and (\ref {eq16}).
The Hamiltonian for intershell rotation taking into account
kinetic energy can be  written in the form
\begin {equation}
\label {eq17}
	H=\frac{1}{2} N_1R_1^2{\dot{\varphi }}^2 (1+\frac {N_1R_1^2}{N_2R_2^2})
	+\frac{1}{2}U_{\star}(1-\cos{(\frac{2\pi\varphi}{\varphi_{\star }})}).
\end {equation}
For clusters with more than two shells, we can only propose
phenomenologic generalization to expression (\ref {eq17}).
Let us label \{$\vec {A}_i; i=1,...,N$\} the set of
eigenvectors corresponding to intershell rotation. Then
to first approximation the Hamiltonian for
intershell rotation becomes
\begin {equation}
\label {eq18}
	H=\frac{1}{2} R_{\star}^2{\dot{\varphi}}^2
	+\frac{1}{2} U_{\star}(1-cos(\frac{2\pi \varphi}{\varphi _{\star}})),
\end {equation}
with
\begin{equation}
\label {eq59}
	R_{\star}^{-1} = \frac{1}{M} \sum_{i=1}^{M} [\vec{r_i}
	\times \vec {A}_i]/r_i^2 \ ,
\end{equation}
where the summation is carried out over the particles of the shell with the
maximum angular velocity.
The value of the parameter $R_{\star}$ is also given in Table I.
Once we have the
Hamiltonian it is not difficult to find the connection between the
barrier angular value $\varphi_{\star}$ and the characteristic frequency for
intershell rotation
\begin {equation}
\label {eq19}
	U_{\star}=2(\omega_{min}\frac{R_{\star} \varphi_{\star}}{2\pi})^2
	=2(\omega_{min} \delta)^2 \ .
\end {equation}
The parameter $\delta$ has a clear physical meaning:
it is the length which a particle travels within a shell when it moves over
the angle $\varphi_{\star}$. The approximate expression
(\ref {eq19}) is shown in Fig. 8 by the solid curve together with
the results of our simulation which are given by the symbols.
Notice that Eq.(\ref{eq19}) describes our numerical results very well over an
energy barrier height variation of more than 8 orders of magnitude.

\section{Density of states and vortex excitations}

	From Fig. 2 we notice that the maximum frequency in the excitation
spectrum, on the average, slowly increases with increasing number of particles.
We can easily  explain this with the aid of the  theory  of  an
infinite system. As it follows from our calculations, and has been
mentioned in previous work \cite {5},
the minimal interparticle distance decreases slowly with the
growth of the cluster size due to the compression of the
inner shell by particles placed at the periphery of the cluster.
As a consequence, the maximum value of the wave vector $k\approx \pi/l_0$
($l_0$ is the mean distance between the particles) and also the
wave frequency will increase weakly with the cluster size.

	For large clusters, it is more interesting to consider
the density of states (DOS) of excitations (phonons) which can be
obtained by a summation of the energy levels, displayed in Fig. 2,
over a frequency interval which we took $\delta \omega =\omega _{max}/40$,
where $\omega _{max}$ is the maximal eigenfrequency.
The results for $N=80$ and $N=300$ particles is shown in Fig. 9.
A characteristic feature in the DOS for all clusters
is the occurrence of two broad maxima.
{}From earlier investigations \cite {9} of classical infinite 2D
systems we know that there are two types of waves  in a 2D Wigner crystal:
the lateral sound waves with dispersion relation $\omega \sim k$
and the longitudinal plasma wave with $\omega \sim \sqrt{k}$, in
the long-wavelength limit.
Using an analytical approximation for the frequency of sound
$\omega _1 ^2 \approx 0.00363\omega _p^2 k^2l_0^2$ and the plasma
frequency $\omega _2^2 \approx \omega _p^2 kl_0(1-0.181483kl_0)$,
taken from \cite {9}, it is possible to show that the
positions of the broad maxima in Fig. 9 are in qualitative agreement with
the ones for an infinite crystal.
In our dimensionless units $\omega _p=2\pi/\rho l_0$.
To obtain the value of $\omega _p$
we used the average particle density $\rho =N/\pi R_o^2$,
where $R_o$ is the radius of the most outer shell. The maximum frequency
of plasma like waves $\omega _{2,max}\approx 1.17\omega _p$ for the
cluster with $N=80$ equals $4.67$ and for $N=300$ is about $5.77$.
Let us assume that the maximum frequency for sound waves is achieved at
$kl_0=\pi $. Then for $N=80$ we obtain $\omega _{1,max}\approx 2.38$ and for
$N=300$ we find $\omega _{1,max}\approx 2.94$ which are slightly larger
than the position of the first maximum appearing in Fig. 9.

	From continuum elastic theory, a 2D electron crystal can be
considered as incompressible at low frequencies \cite {10}.
In order to check if this is still the case for the present finite system
we consider the z-component of the rotor $\Psi_r(k) = \vec{e}_z \cdot
rot \Psi(k)$ and the
divergence $\Psi_d(k) = div \Psi(k)$ of the field of eigenvectors of mode $k$
\begin{mathletters}
\label{eq:all}
\begin{equation}
	\Psi_d(k)= \frac{1}{N} \sum_{i=1}^{N} \psi_{d,i}^2(k) ,
		\label{eq:a}
\end{equation}
\begin{equation}
	\Psi_r(k)= \frac{1}{N} \sum_{i=1}^{N} \psi_{r,i}^2(k) \ .
		\label{eq:b}
\end{equation}
\end{mathletters}
The values $\psi_{d,i}(k)$, and $\psi_{r,i}(k)$ for the $i^{th}$ particle
are given by
\begin{mathletters}
\label{eq2:all}
\begin{equation}
	\psi_{d,i}(k)=\sum_{m=1}^M \left[(\vec{r}_i - \vec{r})\cdot
		\vec{A}_i(k) +
                  (\vec{r}_m - \vec{r})\cdot \vec{A}_m(k) \right]/
		  \mid \vec{r}_i - \vec{r}_m \mid \ , \\
\label{eq2:a} \end{equation}
\begin{equation}
	\psi_{r,i}(k)= \sum_{m=1}^M \left[(\vec{r}_i - \vec{r})\times
		\vec{A}_i(k) +
                  (\vec{r}_m - \vec{r})\times \vec{A}_m(k) \right]/
		  \mid \vec{r}_i - \vec{r}_m \mid \ ,
\label{eq2:b} \end{equation}
\end{mathletters}
where $\vec{r}_m$ are the coordinates of the neighbor particles
and $\vec{A}_i(k)$ is the eigenvector of particle $i$ for mode $k$.
The rotor and divergence of the eigenvector field are shown in Fig. 10 as
function of the excitation frequency for clusters of size $N=80$ and $N=300$.
Notice that for small values of the frequency the rotor of the field of
eigenvectors is larger than the divergence.
As a consequence, in a finite system but with $N$ sufficiently large,
the system is incompressible and one can expect that the low
frequency excitation consists of vortex motion in which the particle
density is not disturbed.
{}From our computer calculations we found that for $N=151$
the minimum eigenfrequency corresponds indeed to the formation of
a vortex/anti-vortex structure (Fig. 5).
Since the total angular moment has to be equal to zero,
those vortexes always come into pairs. With higher
eigenvalues, the number of vortexes rises, although this function is not
necessarily monotonic (see Fig. 5).
Thus when $N$ is sufficiently large the cluster of charged particles can
be described as a viscious non-compressible fluid in case of small wave
vectors. Vortex motion is only expected for sufficiently large $N$ because
the velocity of dissipation of the vortex energy is inversely
proportional to $R^2$, where $R$ is the characteristic radius,
which increases with increasing $N$.

\section {Melting temperatures }

	In Ref. \cite{5} it was shown that the $T=0$ ordered state of the
cluster is destroyed with increasing temperature ($T$). The melting
temperature for this order-disorder transition was obtained by investigating
the radial displacement, the relative intrashell and intershell angular
displacements as function of temperature. Here we will start from the
excitation spectrum of the zero temperature ordered state in order to
calculate the melting temperature using the Lindeman melting criterion
\cite{46}. In the harmonic approximation the mean
square  displacement  is   given by the   following
expression
\begin {equation}
\label {eq12}
	<u^2>=\frac {T}{N} \sum_{k=2}^{2N} \omega_k^{-2} \ ,
\end {equation}
from which we find the melting temperature $T$ using the
relation $<u^2>=\gamma l_0^2$, where $\gamma=0.10$
for a 2D Wigner crystal \cite {22,23}, and $l_0$ is the mean
interparticle distance.
As discussed in previous section, there exists a number of
configurations with $\omega _{min}$ very small which will give
a very large  contribution to the sum (\ref {eq12}).
In order to see what the effect is of these very low frequency excitations
on the melting temperature
we also considered the sum without the first term.
Then we will find the temperature for intershell diffusion.
Because for large clusters, the value of the
interparticle distance around the center and near the
periphery can be considerably different, therefore
we will use the mean value of relative
displacement in order to define the melting temperature
\begin {equation}
\label {eq13}
	T=\gamma N \left [\sum _{i=1}^{N} l_i^{-2}
	\sum_{k=2}^{2N} \vec {A}_i^2(k)/ \omega_k^2 \right]^{-1} \ ,
\end {equation}
where $\vec{A}_i(k)$ is the displacement vector for the $i$-particle in
mode $k$, and $l_i$ is the mean interparticle distance for the $i$th particle.

	The numerical results are shown in Fig. 11. As we expect there
is a significant difference in the transition temperature whether
intershell rotation is taken into account or not. These results agree
qualitatively with the results of Ref. \cite{5} where it was found that:
i) for clusters with a small number of particles the angular order is
destroyed at much lower temperatures than the radial order which agrees
with the large difference in melting temperatures shown in Fig. 11;
ii) for larger clusters both temperatures are practically equal as is
also apparent in Fig. 11. Orientational order and radial order disappear
practically at the same temperature for $N>40$; and iii) the melting
temperature
at which intershell diffusion sets in, is a decreasing function of the number
of particles in the cluster up to about $N \sim 20 \div 40$ beyond which it
starts to increase, which agrees qualitatively with Fig. 11. The magnitude
of the transition temperature found in the present approach is slightly
higher than found in the Monte Carlo study of Ref. \cite{5}. This is
a consequence of the present harmonic approximation which has a limited
validity near the melting temperature.
The melting temperature for intershell rotation (top part of Fig. 11) is
strongly influenced by the value of $\omega_{min}$ which is proportional
to the rigidity of the cluster agains intershell rotations. In fact it is
a measure of the stability of the cluster against intershell rotations.
As was mentioned before the value of $\omega_{min}$, and also $U_{\star}$,
is determined by the configuration of the cluster. Clusters with a
magic number of particles have a large melting temperature for intershell
rotation. These fine details were not present in Ref. \cite{5}.

	It is known that for an
infinite crystal the sum $(\ref {eq13})$ diverges logarithmically in the
low-wavelength which is due to the presence of
lateral sound waves. Therefore one uses the average
square displacement of interparticle distance in Lindeman's melting
criterion. In our case, such criterion gives the relation
\begin {equation}
\label {eq14}
	T=\gamma N\left[ \sum_{i=1}^{N} l_i^{-2} \sum_{k=2}^{2N} \omega_k^{-2}
	\frac{1}{M} \sum_{m=1}^{M} (\vec{A}_i(k)-\vec{A}_m(k))^2 \right]^{-1}
\end {equation}
where the sum over $m$ runs over the M neighbor particles.
The numerical results obtained using Eq.(\ref{eq14}) is shown in Fig. 12.
These results are very close to those found in Ref. \cite{5} with the
exception that here near $N \sim 150$ a maximum is found while the
Monte Carlo results slowly increases towards the $N \rightarrow \infty$
value. We want to emphasize that if the number of particles is not too
large, the transition
temperature obtained with the second criterion (\ref {eq14}) is
lower than the one from Eq. (\ref {eq13}).
This indicates that the particles  mainly move towards each other,
and only for $N\geq200$, the effect of small wave vectors
begin to appear. In the latter case
the neighbor particles move with the same velocity and the
difference in the value of critical  temperatures obtained using the
spectrum of the eigen vibrations (Fig. 11)
and the Monte-Carlo technique \cite {5} is very small.

\section {Specific heat}

	Before we already mentioned, that in order to obtain the ground state we
generated many initial configurations and in so doing not only stable
states but also metastable states were obtained.
Thus, if we also calculate the spectrum of normal modes $\omega_{k}$
for each local minimum we can easily obtain the partition
function within the harmonic approximation and consequently all the
thermodynamic quantities like the free energy, the specific heat,...
Such an approach was followed in Refs. \cite{19} and \cite{20} for 3D clusters
where also the influence of anharmonicity and saddle
points on the partition function was studied.
In Refs. \cite{19,20} only, the main
characteristics of the spectrum of normal modes was used.
Here we know the complete spectrum of our finite 2D system and
are therefore able to calculate the partition function more correctly.

 	In the quasiclassical approximation the partition function for
a cluster with N particles is given by \cite {21}
\begin{equation}
	Z(T)=(2\pi \hbar )^{-2N} \int
	d\vec {q}d\vec{p} \exp{(-H(\vec{q},\vec{p})/k_bT)}\ ,
\end{equation}
where $\vec{q}= (\vec{r}_1,...,\vec{r}_N)$,
$\vec{p} = (\vec{p}_1,...,\vec{p}_N)$ are $2N$ dimensional vectors.
The partition function can be written as
\begin {equation}
\label{eq20}
	Z(T)=\sum _{m=1}^{M}\exp {(-U_m/T)}Z_m(T) ,
\end {equation}
where $Z_m$ is the partition function of the
$m$th metastable state whose energy differs with the ground state
by an amount $U_m$. The dimensionless units for temperature and specific
heat are used here and below. In the
vicinity of this $m$th metastable state, the Hamiltonian is quadratic in
the normal coordinates.
Because the energy barrier for intershell rotation is small, the
effect of anharmonicity will already appear at low temperatures.
Therefore we will integrate only over a small region of
particle motion $\mid q_i\mid \leq \sqrt {2U_m(k)/T\omega_{k,m}}$
which results in
\begin {equation}
\label {eq21}
	Z_m=g_mZ_{rot}
	\prod_{k=2}^{2N} \frac {T} {\hbar \omega_{k,m}}
	erf(\sqrt{U_m(k)/T} ),
\end {equation}
where $Z_{rot}\propto \sqrt {T}$ is the part of the partition function
resulting from the rotational degrees of freedom,
$g_m=2\pi/\theta_{\star}$ is the degeneracy of the $m$th state, which is
determined by the number of particles occupying a shell,
$U_m(k)$ is the  barrier height for
normal mode $k$, and $erf(x)$ is the error function.
These parameters are given in Table II for a number of metastable states.

	For convenience let us consider only one normal mode.
At low temperature $T\ll U_m(k)$
expression (\ref {eq21}) results in the usual value for the
specific heat for a harmonic oscillator $C=1$.
For high temperature $T\ll U_m(k)$ the specific heat equals $1/2$ as for
free motion. For the intermediate temperature region $T\sim U_m(k)$ expression
(\ref {eq21}) gives an interpolation between these two limiting cases.
Unfortunately, we know only the value of the barrier
height for intershell rotation. For the remaining normal modes,
we will use the analogy
with the Lindeman criterion to write the  phenomenological relation
\begin{equation}
	U_m(k)=\gamma_u N \frac {\omega_{k,m}^2l_0^2}{2}\ ,
\end{equation}
where $l_0$ is the mean interparticle  distance, and $\gamma _u
=0.2\div 0.3 $.
The above expression is then used in the numerical evaluation of the
partition function (\ref{eq21}) and (\ref{eq20}).
Below we will mainly deal with  the specific heat
\begin{equation}
	C=\frac {\partial}{\partial T} T^2 \frac {\partial \log
	{Z}}{\partial T} \ ,
\end{equation}
which is shown by the solid curve in Fig. 13.

 	General features of the behavior of the specific heat as function of
$T$ and $N$ can be predicted
without detailed information regarding the metastable states and the
spectrum of the normal modes. At low temperatures
such that $T\ll (U_m(k),\: U_m)$
the specific heat is only determined by the ground state, and the effect of
anharmonicity is not essential. Consequently $C=2N-1/2$ as is apparent
in Fig. 13..
Usually the barrier for intershell rotation is the smallest energy,
which is also smaller than the difference in energy between
the ground state and the metastable states.
In the temperature range
$U_1(k=2) \ll T \ll (U_1(k\ne 2),\: U_{m}) $
the specific heat will be constant and having the value $C=2N-1$.
Such a small reduction in $C$ is visible in Fig. 13 near $T \sim 10^{-3}$.
With further increase of the temperature, the behavior
of the specific heat is determined by
the competition of two processes. On the one hand, transitions to
metastable states which lead to an increase
of the specific heat, and on the other hand,
the effect of anharmonicity which will reduce $C$. This interplay will
lead to peaks in the specific heat as is apparent in Fig. 13.
Note that the position of the peak does not equal the melting temperature.

	In the order-disorder transition region the applicability of the above
approach is  questionable. Therefore, we also calculated the specific heat
using the standard Monte-Carlo technique.
As the initial state we took the ground state of our system.
Then we fix the temperature and execute
$10^5$ steps of the Metropolis algorithm to allow the  system
to achieve equilibrium. Next about $(4\div 10)\cdot 10^6$ steps
of the Metropolis algorithm are made
in order to reduce the statistical error.
The specific heat is then found using the following formula
\begin{equation}
	C=N+(<E^2>-<E>^2)/T^2\ ,
\end{equation}
where $E$ is the potential energy for the system with  $N$  particles.
In Fig. 13 we compare the results from
the Monte-Carlo simulation (full dots) with the above
results (full curve) which are based on the excitation spectrum of the $T=0$
stable and metastable states. Note that for the small cluster with $N=9$
very good agreement is obtained. For the other two clusters good
quantitative agreement is found at low temperature while at intermediate
and high temperatures only the qualitative behavior is correctly described.
Thus for large clusters the
approximate model is not able to give a satisfactory description
of the effect of anharmonicity. Nevertheless there is
qualitative agreement in the position of the maxima.
We have tried to vary the parameter $\gamma _u$ and to change
the integration interval for allowed particle motion in Eq.(26)
but we were not able to obtain any better agreement.

\section {Conclusion}

	We have presented the results of a numerical simulation of
the ground state and the spectrum of normal modes of classical 2D clusters
with quadratic confinement. The barriers for
intershell rotation and the specific heat are also obtained.
The Lindeman melting criterion in conjunction with the $T=0$ excitation
spectrum of the ground state configuration was used to obtain the
order-disorder
transition temperatures for angular and radial melting.

	For systems with axial symmetry, and an intermediate number of
particles the normal mode with the lowest frequency corresponds
to intershell rotation if there are at least two shells.
A low excitation energy for intershell rotation is found for clusters
which have a shell configuration such that the number of particles on each
shell
have no common multiple. If the number of particles in the outer
shell is an integer multiple of the number of particles in the inner shell,
the cluster will be most stable against intershell rotation which define
the clusters with magic numbers. Such clusters also have a large melting
temperature for intershell rotation.
Distortion of the axial symmetry of the external potential,
will lead to a rise in the eigenfrequency and in the barrier height for
intershell rotation. For large clusters, i.e.
$N>100$, the normal mode with the lowest frequency corresponds
to a vortex/anti-vortex excitation.

\section {Acknowledgments }

	We wish to thank our colleague V.M. Bedanov for fruitful discussions.
Part of this work is supported by INTAS-93-1495,
the Human Capital and Mobility network programme No. ERBCHRXT 930374 and
the Belgian National Science Foundation.

\newpage
\begin{figure}
\caption{Accuracy of the calculated groundstate energy versus
the number of simulation steps using the Monte-Carlo technique and
the present optimized version of the Newton
technique for a cluster consisting of $N=13$ particles.}
\end{figure}

\begin{figure}
\caption{Excitation spectrum of normal modes as function of the number of
particles in the cluster. The frequency is in units of $\omega/\sqrt 2$.}
\end{figure}

\begin{figure}
\caption{Eigenvectors for a cluster with $N=9$ particles
for different mode number $k$
which correspond to the eigenfrequencies $\omega _1=0$,
$\omega _2\approx 0.127$, $\omega _7 \approx 1.414$,
$\omega _{15} \approx 2.449$.}
\end{figure}

\begin{figure}
\caption{Eigenvectors corresponding to the minimal frequency for clusters with
$N=19, 20, 39$, and $60$ particles.}
\end{figure}

\begin{figure}
\caption{Eigenvectors for the cluster with $N=151$ particles for
four different values of the mode number $k$.}
\end{figure}

\begin{figure}
\caption{Trajectories of the particles for intershell rotational motion in
case of clusters of $N=9, 19, 20$ and $38$ particles.}
\end{figure}

\begin{figure}
\caption{Cluster energy versus shell turn angle for the shell with
the maximal angular velocity for clusters with $N=9$, and $40$ particles.}
\end{figure}

\begin{figure}
\caption{Energy barrier of intershell rotation versus $(\omega \delta)^2$ where
$\omega =\omega_{min}$ is the frequency for intershell rotation and
$\delta$ is the linear distance traveled by a particle over one
angular period.}
\end{figure}

\begin{figure}
\caption{Density of phonon states for clusters with $N=80$ and $300$
particles.}
\end{figure}

\begin{figure}
\caption{The value of the divergence (a,c) and rotor (b,d) of
the displacement field for clusters with $N=80$ (a,b) and $N=300$ (c,d)
particles.}
\end{figure}

\begin{figure}
\caption{Temperature of cluster melting obtained using the
Lindeman criterion with and without intershell rotation.}
\end{figure}

\begin{figure}
\caption{Melting temperature for large clusters as obtained from
Lindeman's melting criterion excluding intershell rotation and
incorporating the relative displacement of neighbor particles.}
\end{figure}

\begin{figure}
\caption{Specific heat for clusters with $N=9, 25$ and $34$ particles
as function of temperature.
Solid lines are the results as obtained from an approximate calculation
of the partition function and the solid dots are results from the
Monte-Carlo simulation.}
\end{figure}

\newpage
\begin {center} { TABLE I} \end {center}

%TableI. Ground-state configuration for the system with
%parabolic-confinement potential.
Table I. Shell configuration ($N_1,N_2,\ldots $) for clusters with
N-particles with parabolic confinement.
The minimal excitation frequency ($\omega _{min}$ in units of
$\omega_o /\sqrt{2}$),
the period ($\varphi _{\star }$) in degree units and the barrier height
($U_{\star }$) for intershell rotation are given together with the
parameter $R_{\star }$ for the ground state of the cluster.
\begin {center}
\begin {tabular}{llcrlr} \hline
\hspace {1 cm} $N$\hspace {0.5cm}
&\hspace {1 cm} $N_1,N_2,\ldots$\hspace {0.5cm}
&\hspace {1 cm} $\omega _{min}$\hspace {0.5cm}
&\hspace {1.5cm} $\varphi _{\star }$\hspace {0.1cm}
&\hspace {1.5cm} $U_{\star }$\hspace {0.1cm}
&\hspace {1.0cm} $R_{\star }$ \hspace {0.5 cm}\\ \hline
\hspace {1 cm} $9$  & $2,7$               & $1.268\times10^{-1}$ & $24.9$
&\hspace {1 cm} $8.44\times10^{-5}$ & $1.349 $  \\
\hspace {1 cm} $10$ & $2,8$                & $8.910\times10^{-2}$ & $43.9$
&\hspace {1 cm} $1.20\times10^{-4}$ & $1.427 $   \\
\hspace {1 cm} $11$ & $3,8$               & $2.451\times10^{-2}$ & $14.2$
&\hspace {1 cm} $2.42\times10^{-6}$ & $0.882$   \\
\hspace {1 cm} $12$ & $3,9$               & $5.308\times10^{-1}$ & $38.3$
&\hspace {1 cm} $7.33\times10^{-3}$ & $0.912$   \\
\hspace {1 cm} $13$ & $4,9$               & $6.002\times10^{-4}$ & $ 9.7$
&\hspace {1 cm} $1.06\times10^{-9}$ & $0.672$   \\
\hspace {1 cm} $14$ & $4,10$              & $4.940\times10^{-2}$ & $16.8$
&\hspace {1 cm} $2.60\times10^{-5}$ & $0.644$   \\
\hspace {1 cm} $15$ & $5,10$              & $4.599\times10^{-1}$ & $32.8$
&\hspace {1 cm} $1.11\times10^{-2}$ & $0.564$   \\
\hspace {1 cm} $16$ & $1,5,10$            & $4.924\times10^{-1}$ & $31.9$
&\hspace {1 cm} $2.03\times10^{-2}$ & $0.449$   \\
\hspace {1 cm} $17$ & $1,6,10$            & $5.416\times10^{-2}$ & $10.4$
&\hspace {1 cm} $3.44\times10^{-5}$ & $0.374$   \\
\hspace {1 cm} $18$ & $1,6,11$            & $6.141\times10^{-3}$ & $ 4.8$
&\hspace {1 cm} $1.03\times10^{-7}$ & $0.360$   \\
\hspace {1 cm} $19$ & $1,6,12$            & $6.676\times10^{-1}$ & $26.6$
&\hspace {1 cm} $3.14\times10^{-2}$ & $0.396$   \\
\hspace {1 cm} $20$ & $1,7,12$            & $1.031\times10^{-4}$ & $ 4.0$
&\hspace {1 cm} $2.01\times10^{-11}$ & $0.334$   \\
\hspace {1 cm} $21$ & $1,7,13$            & $3.174\times10^{-3}$ & $ 3.5$
&\hspace {1 cm} $2.18\times10^{-8}$ & $0.294$   \\
\hspace {1 cm} $22$ & $2,8,12$            & $2.934\times10^{-1}$ & $12.5$
&\hspace {1 cm} $5.44\times10^{-3}$ & $0.257$   \\
\hspace {1 cm} $23$ & $2,8,13$            & $1.287\times10^{-1}$ & $11.6$
&\hspace {1 cm} $4.74\times10^{-4}$ & $0.258$   \\
\hspace {1 cm} $24$ & $3,8,13$            & $2.762\times10^{-2}$ & $ 2.7$
&\hspace {1 cm} $5.86\times10^{-6}$ & $0.118$   \\
\hspace {1 cm} $25$ & $3,9,13$            & $1.138\times10^{-1}$ & $ 7.4$
&\hspace {1 cm} $2.47\times10^{-4}$ & $0.210$   \\
\hspace {1 cm} $26$ & $3,9,14$            & $1.041\times10^{-1}$ & $ 7.0$
&\hspace {1 cm} $1.81\times10^{-4}$ & $0.212$   \\
\hspace {1 cm} $27$ & $4,9,14$            & $1.311\times10^{-2}$ & $12.0$
&\hspace {1 cm} $1.04\times10^{-6}$ & $0.611$   \\
\hspace {1 cm} $28$ & $4,10,14$           & $5.682\times10^{-2}$ & $ 5.6$
&\hspace {1 cm} $8.31\times10^{-5}$ & $0.179$   \\
\hspace {1 cm} $29$ & $4,10,15$           & $3.911\times10^{-2}$ & $ 9.5$
&\hspace {1 cm} $1.58\times10^{-4}$ & $0.183$   \\
\hspace {1 cm} $30$ & $5,10,15$           & $2.974\times10^{-1}$ & $18.6$
&\hspace {1 cm} $1.47\times10^{-2}$ & $0.172$   \\
\hspace {1 cm} $31$ & $5,11,15$           & $2.351\times10^{-2}$ & $ 5.1$
&\hspace {1 cm} $1.52\times10^{-5}$ & $0.167$   \\
\hspace {1 cm} $32$ & $1,5,11,15$         & $2.971\times10^{-2}$ & $ 5.1$
&\hspace {1 cm} $2.57\times10^{-5}$ & $0.137$   \\
\hspace {1 cm} $33$ & $1,6,11,15$         & $6.805\times10^{-2}$ & $10.0$
&\hspace {1 cm} $1.15\times10^{-4}$ & $0.257$   \\
\hspace {1 cm} $34$ & $1,6,12,15$         & $2.379\times10^{-1}$ & $ 8.6$
&\hspace {1 cm} $3.68\times10^{-3}$ & $0.131$   \\
\hspace {1 cm} $35$ & $1,6,12,16$         & $6.585\times10^{-2}$ & $ 5.5$
&\hspace {1 cm} $1.12\times10^{-4}$ & $0.134$   \\
\hspace {1 cm} $36$ & $1,6,12,17$         & $8.963\times10^{-3}$ & $ 2.6$
&\hspace {1 cm} $5.04\times10^{-7}$ & $0.131$   \\
\hspace {1 cm} $37$ & $1,7,12,17$         & $3.214\times10^{-3}$ & $ 5.3$
&\hspace {1 cm} $4.66\times10^{-8}$ & $0.317$   \\
\hspace {1 cm} $38$ & $1,7,13,17$         & $6.134\times10^{-3}$ & $ 4.8$
&\hspace {1 cm} $3.28\times10^{-6}$ & $0.113$   \\
\hspace {1 cm} $39$ & $2,8,12,17$         & $2.231\times10^{-1}$ &    -
&\hspace {1 cm}      -              & $0.091$   \\
\hspace {1 cm} $40$ & $4,6,13,17$         & $1.242\times10^{-1}$ & $14.8$
&\hspace {1 cm} $2.21\times10^{-3}$ & $0.118$   \\
\hspace {1 cm} $41$ & $4,6,14,17$         & $1.237\times10^{-1}$ & $ 4.8$
&\hspace {1 cm} $5.90\times10^{-4}$ & $0.092$   \\
\hspace {1 cm} $42$ & $3,8,14,17$         & $3.340\times10^{-2}$ &    -
&\hspace {1 cm}       -             & $0.096$   \\
\hspace {1 cm} $43$ & $3,9,14,17$         & $5.010\times10^{-2}$ & $ 7.0$
&\hspace {1 cm} $2.97\times10^{-4}$ & $0.200$   \\
\hspace {1 cm} $44$ & $3,9,14,18$         & $1.552\times10^{-1}$ & $ 7.5$
&\hspace {1 cm} $3.40\times10^{-3}$ & $0.092$   \\
\hspace {1 cm} $45$ & $3,9,15,18$         & $1.962\times10^{-1}$ & $13.4$
&\hspace {1 cm} $2.66\times10^{-2}$ & $0.094$   \\
\hspace {1 cm} $46$ & $3,9,15,19$         & $8.425\times10^{-1}$ & $ 4.3$
&\hspace {1 cm} $2.40\times10^{-4}$ & $0.093$   \\
\hspace {1 cm} $47$ & $4,10,15,18$        & $1.850\times10^{-1}$ &     -
&\hspace {1 cm}       -             & $0.034$   \\
\hspace {1 cm} $48$ & $4,10,15,19$         & $1.242\times10^{-1}$ & $ 9.1$
&\hspace {1 cm} $2.37\times10^{-3}$ & $0.130$   \\
\hspace {1 cm} $49$ & $4,10,16,19$         & $1.511\times10^{-1}$ & $ 6.2$
&\hspace {1 cm} $2.05\times10^{-3}$ & $0.079$   \\
\hspace {1 cm} $50$ & $4,10,16,20$         & $7.535\times10^{-2}$ & $12.0$
&\hspace {1 cm} $3.32\times10^{-3}$ & $0.088$   \\
\hspace {1 cm} $51$ & $5,11,16,19$         & $7.530\times10^{-2}$ &    -
&\hspace {1 cm}       -            & $0.189$   \\
\hspace {1 cm}      &                      &                      &
&                    &           \\
\hspace {1 cm} $60$ & $1,7,13,18,21$       & $7.420\times10^{-2}$ & $ 7.0$
&\hspace {1 cm} $7.11\times10^{-4}$ & $0.059$   \\
\hspace {1 cm} $70$ & $6,6,15,20,23$       & $1.220\times10^{-2}$ &    -
&\hspace {1 cm}       -            & $0.097$   \\
\hspace {1 cm} $80$ & $1,6,12,17,22,22$    & $1.840\times10^{-2}$ & $ 4.9$
&\hspace {1 cm} $1.19\times10^{-5}$ & $0.117$   \\ \hline
\end {tabular}
\end {center}

\newpage
\begin {center} { TABLE II} \end {center}

Table II. Shell configuration ($N_1,N_2,\ldots $) for some metastable states
for a number of different clusters. $U_m$ is the energy difference of the
metastable configuration with the ground state energy and
$W_m=\prod {\omega_{k,m=1}}/\prod {\omega _{k,m}}$ is the
relative statistical weight.
\begin {center}
\begin {tabular}{llcr} \hline
\hspace {1 cm} $N$\hspace {0.5cm}
&\hspace {0.1cm} $N_1,N_2,\ldots$\hspace {1.0cm}
&\hspace {0.7cm} $U_m$\hspace {1.0cm}
&\hspace {1.3 cm} $W_m$\hspace {0.2cm}\hspace {0.5 cm}\\ \hline
%\hspace {1 cm} $9$   & $2,7       $  & $0$                & $1$    \\
\hspace {1 cm}  $9$    & $1,8       $  & $5.526\times10^{-2}$& $3.14$ \\
%\hspace {1 cm}        & $3,9,13    $  & $0$                & $1$    \\
\hspace {1 cm}  $25$    & $3,8,14    $  & $5.308\times10^{-2}$& $0.81$ \\
\hspace {1 cm}   -   & $4,8,13    $  & $1.013\times10^{-1}$& $22.70$ \\
%\hspace {1 cm} $34$  & $1,6,12,15 $  & $0$                & $1$    \\
\hspace {1 cm}  $34$ & $1,6,11,16 $  & $9.114\times10^{-3}$& $15.91$ \\
\hspace {1 cm}   -    & $1,7,11,15 $  & $1.003\times10^{-2}$& $79.70$ \\
\hspace {1 cm}   -   & $1,5,11,17 $  & $1.581\times10^{-2}$& $ 7.22$ \\
\hspace {1 cm}   -   & $6,12,16   $  & $1.847\times10^{-2}$& $60.60$ \\ \hline
\end {tabular}
\end {center}

\begin{thebibliography}{BM}
\bibitem[*]{f:gnu}
To whom correspondence should be addressed.
Electronic mail: peeters@ nats.uia.ac.be
\end{thebibliography}
\end{document}